\documentclass[12pt] {article}
\usepackage{latexsym}{\rm }
\usepackage[dvips]{graphics,color}
\usepackage{graphpap}
\usepackage{alltt}
\usepackage{graphicx}
\usepackage{color}

\newcommand{\beq}{\begin{equation}}
\newcommand{\feq}[1]{\label{#1} \end{equation}}
\newcommand{\beqr}{\begin{eqnarray}}
\newcommand{\feqr}{\end{eqnarray}}
\def\non{\nonumber}

\newcommand{\rf}[1]{(\ref{#1})}

\definecolor{red}{rgb}{1,0,0}
\begin{document}

\begin{center}


{\Large \bf A Diagrammatic Approach for the Coefficients of the Characteristic Polynomial}\\
[4mm]

\large{Agapitos Hatzinikitas} \\ [5mm]

{\small University of Aegean, \\
School of Sciences, \\
Department of Mathematics, \\
Karlovasi, 83200\\
Samos Greece \\
Email: ahatz@aegean.gr}\\ [5mm]

\end{center}
\begin{abstract}
{\bf In this work we provide a novel approach for computing the coefficients
of the characteristic polynomial of a square matrix. We demonstrate that
each coefficient can be efficiently represented by a set of circle graphs.
Thus, one can employ a diagrammatic approach to determine the coefficients
of the characteristic polynomial.}
\end{abstract}
\section{Introduction}
A variety of different branches in Mathematical Physics boil down to calculating the eigenvalues of an $n\times n$ matrix $A$ over some field $K$ usually taken to be the real line or the complex plane. When we are interested in all the latent roots of the characteristic polynomial of a matrix and not in its latent vectors we usually expand out the expression \cite{f1} 
\beqr
f(x)&=& det(A-xI)=\prod_{i=1}^{n}(x-\lambda_i) \non \\
&=& x^n+(-1)^{1}e_1(\lambda)x^{n-1}+(-1)^{2} e_2(\lambda)x^{n-2}+(-1)^{3}e_3(\lambda)x^{n-3}+\cdots +(-1)^n e_n(\lambda) \non \\
\label{a1}
\feqr 
where $e_k(\lambda)=\sum_{1\leq i_1<\cdots<i_k\leq n}\lambda_{i_1}\cdots\lambda_{i_k}$ are the elementary symmetric polynomials in $k$ eigenvalues $\lambda_{i_s}, \, s=1,\cdots, k$. There are different methods to determine explicitly $e_k(\lambda)$ in terms of a sum of trace products $\prod_{r=1}^l \left(tr(A^{r})\right)^{m_r}$, over all possible partitions $(1^{m_1},\cdots,l^{m_l})$ \footnote{We adopt the following notation: $n=(1^{m_1},2^{m_2}, \cdots, r^{m_r}, \cdots)$ where $m_i$ counts the number of parts of $n$ which are equal to $i$. It is called the multiplicity of $i$ in $n$.} of the positive integer $k$. Undoubtably, all methods for high order matrices become very tedious. 
\par One method for an $n\times n$ matrix is based upon the multinomial formula
\beqr
(\lambda_1+\cdots +\lambda_n)^m=\sum_{k_1,\cdots,k_n}\left(\begin{array}{c} m\\ k_1,\cdots,k_n\end{array}\right)\lambda_1^{k_1}\cdots \lambda_n^{k_n}
\label{a11}
\feqr 
where the summation is taken over all sequences of nonnegative integer indices $k_1,\cdots,k_n$ such that $\sum_{i=1}^n k_i=m$ and the cofficients are given by
\beqr
\left(\begin{array}{c} m\\ k_1,\cdots,k_n\end{array}\right)=\frac{m!}{k_1!\cdots k_n!}.
\label{a12}
\feqr 
For $n=2$ relation \rf{a11} gives   
\beqr
\left(\sum_{i_1=1}^{n}\lambda_{i_1}\right)^2 &=& \sum_{i_1=1}^{n}\lambda_{i_1}^2 +2\sum_{1\leq i_1<i_2\leq n} \lambda_{i_1}\lambda_{i_2} 
\label{a2}
\feqr
from which we find 
\beqr
e_2(\lambda)&=& -\frac{1}{2}\left((tr(A^2)-(trA)^2)\right). 
\label{a21}
\feqr
Similarly for $e_3(\lambda)$ we have
\beqr
e_3(\lambda)&=& \frac{1}{6}\left(\left(\sum_{i_1=1}^n \lambda_{i_1}\right)^3-2\sum_{i_1=1}^{n}\lambda_{i_1}^3-3\sum_{i_1\neq i_2=1}^n\lambda_{i_1}^2\lambda_{i_2}\right) \non \\
&=&-\frac{1}{6}\left(2tr(A^3)+3tr(A^2)trA-(trA)^3\right).
\label{a3}
\feqr
Note that the three terms in \rf{a3} correspond to the partitions $3^1=(1^1,2^1)=1^3$. In the same vein we can compute any $e_k(\lambda)$. 
\par A second method is to use $Newton-Girard's$ formulas which give the connection between the coefficients $p_r(\lambda)=(-1)^{-r}e_{r}(\lambda)$ and the power sums
\beqr
S_r&=&\sum_{i_1=1}^{k}\lambda_{i_1}^r, \non \\
S_r+S_{r-1}p_1+&\cdots& + S_1p_{r-1}+rp_r=0, \quad r\leq n.
\label{a4}
\feqr 
From \rf{a4} it follows that
\beqr
p_1&=& -S_1 \non \\
p_2&=& -\frac{1}{2}(S_1p_1+S_2) \non \\
&\cdots& \non \\
p_r&=&-\frac{1}{r}(S_1p_{r-1}+\cdots + S_{r-1}p_1+S_r).
\label{a5}
\feqr
Relations \rf{a5} determine the coefficients $p_r$ and the process is called $Le \, Verrier's$ method. 
\section{The New Method}

The method we propose for the evaluation of the coefficients of the charactristc polynomial is based upon the knowledge of the number of partitions $q(k)$ for the index $k$ which specifies the elementary symmetric polynomial $e_k(\lambda)$. The generating function of $q(k)$ is given by Euler's formula \cite{f2}
\beqr
F_{tot}(x)&=& \frac{1}{\prod_{k=1}^{\infty}(1-x^k)}=\sum_{k=0}^{\infty}q(k)x^k. 
\label{a10}
\feqr
Other useful generating functions are 
\beqr
F_{even}(x)=\frac{1}{\prod_{k=1}^{\infty}(1-x^{2k})}, \quad F_{odd}(x)=\frac{1}{\prod_{k=0}^{\infty}(1-x^{2k+1})}, \quad F_{uneq.}(x)=\prod_{k=1}^{\infty}(1+x^k)   
\label{a101}
\feqr
where $F_{even}$, $F_{odd}$ and $F_{uneq.}$ enumerate partitions of $k$ into $even$, $odd$ and $unequal$ parts respectively. 
\par Having the partitions at hand we proceed by imposing certain rules for the construction of diagrams:
\begin{enumerate}
\item[($1$)] Each matrix $(A_{ij})$ is represented by a line segment with indices $i, j$ attached to the endpoints. The trace of $A$ is graphically formed by gluing the two endpoints and thus resulting in a circle graph.
\setlength{\unitlength}{.2in}
\begin{center}
\begin{picture}(1,1)(-1,-1)
\put(0,0){\line(1,0){2}}
\put(-.5,0){\makebox(0,0)[l]{i}}
\put(-4.5,0){\makebox(0,0)[l]{$(A_{ij})\longrightarrow $}}
\put(2.5,0){\makebox(0,0)[r]{j}}
\put(-4.5,-2){\makebox(0,1)[l]{$tr(A)\longrightarrow$}}
\put(1.,-1.5){\circle{1}}
\put(1.,-1.){\circle*{.2}}
\end{picture}
\end{center}
\item[($2$)] For every positive integer $k$ we associate a single circle graph with $k$ points (we call it a \textit{k-circle} from now on) and contributing a factor $-tr(A^k)$.
\setlength{\unitlength}{.2in}
\item[($3$)] A partition of $k=(1^{m_1},2^{m_2}, \cdots, r^{m_r}, \cdots)$ is represented by a set of circle graphs which can be constructed by a cutting and sewing procedure from the k-circle
\setlength{\unitlength}{.2in}
\begin{center}
\begin{picture}(4,2)(-1,-1)
\put(-8.5,0){\makebox(0,0)[l]{$\longrightarrow$}}
\put(-6.5,0){\makebox(0,0)[r]{$\Biggl($}}
\put(-10.5,0){\circle{2.}}
\put(-10.5,1){\circle*{.2}}
\put(-10.5,-1){\circle*{.2}}
\put(-11.5,0){\circle*{.2}}
\put(-11.35,0.5){\circle*{.2}}
\put(-11.35,-0.5){\circle*{.2}}
\put(-9.65,0.5){\circle*{.2}}
\put(-9.65,-0.5){\circle*{.2}}
\put(-9.52,0){\circle*{.2}}
\put(-8.5,-2.){\makebox(0,0)[r]{$k-$circle}}
\put(-6,0){\circle{1}}
\put(-6,.5){\circle*{.2}}
\put(-6.5,0){\makebox(2.5,0)[r]{$\Biggr)^{m_1}$}}
\put(-6.2,0){\makebox(2.7,0)[r]{$+$}}
\put(-5.5,0){\makebox(2.7,0)[r]{$\Biggl($}}
\put(-2.4,0){\circle{1}}
\put(-2.4,.5){\circle*{.2}}
\put(-2.4,-.5){\circle*{.2}}
\put(-3,0){\makebox(2.5,0)[r]{$\Biggr)^{m_2}$}}
\put(-1.7,0){\makebox(2.7,0)[r]{$+ \cdots $}}
\put(-1.,0){\makebox(2.7,0)[r]{$\Biggl($}}
\put(2.,0){\circle{1}}
\put(2.,.5){\circle*{.2}}
\put(2.,-.5){\circle*{.2}}
\put(2.5,0){\circle*{.2}}
\put(1.7,0){\makebox(2.5,0)[r]{$\Biggr)^{m_r}$}}
\put(3.2,0){\makebox(2.5,0)[r]{$+ \cdots$}}
\end{picture}
\end{center}
where the powers on the right handside stand for the multiplicity of each graph.  
\vspace{.2cm}
\item[($4$)] If $r^{m_r}$ is a single partition of $k$ then the coefficient is given by  
\beqr
(-1)^{m_r}tr(A^r)\frac{((r-1)!)^{m_r}}{(m_r)!}\prod_{l=0}^{m_r-1}\left(\begin{array}{c}k-lr\\r\end{array}\right) 
\label{b1}
\feqr
\end{enumerate}
The interpretation of each factor in \rf{b1} is as follows:
\begin{enumerate}
\item[($\alpha$)] The number of circular permutations of $r$ distinct points on the circle is $(r-1)!$. 
\item[($\beta$)] The number of permutations of $r$-circles $m_r$ times is $(m_r)!$. 
\item[($\gamma$)] The ways of extracting $r$ points each time from a $k$-circle, $m_r$ times,  without replacement and disregarding order is given by the product.  
\end{enumerate}
Note that if we sum up the absolute values of factors \rf{b1} for all possible partitions of $k$ then we recover $k!$. 
\par A more involved case study will be the partition $k=(r^{m_r}, s^{m_s})$. The combinatorial factor now reads
\beqr
(-1)^{m_r+m_s}tr(A^r)tr(A^s)\frac{((r-1)!)^{m_r} ((s-1)!)^{m_s}}{(m_r)! (m_s)!} \prod_{l=0}^{m_r-1}\left(\begin{array}{c}k-lr\\r\end{array}\right)\prod_{\rho=0}^{m_s-1}\left(\begin{array}{c}k-rm_r-\rho s\\s\end{array}\right).
\label{b2}
\feqr  
As an application consider the case of $k=6$. The total number of partitions is $q(6)=11$ which splits into $q_{even}(6)=3$ and $q_{odd}(6)=q_{uneq.}(6)=4$ parts. Adopting the convention $T^k_l=(tr(A^l))^k$ a list of the contributions of all partitions of 6 is given in the following table 
\begin{table}[h]\caption{The summands of $e_6$ applying the diagrammatic approach.}
\begin{center}
\begin{tabular}{|c|c|}\hline
Partitions & Coefficients \\
\hline \hline
(6) & $ -120 \, \, T_6$  \\
\hline
(5,1) & $144 \, \,  T_5 T_1$  \\
\hline
(4,2) &  $90 \, \, T_4 T_2$ \\
\hline
$(3^2)$ &  $40 \, \, T_3^2$ \\
\hline
$(4,1^2)$ & $-90 \, \, T_4 T_1^2$ \\
\hline
$(3,2,1)$ & $- 120 \, \, T_3T_2T_1$ \\
\hline
$(2^3)$ &  $-15 \, \, T_2^3$ \\
\hline
$(3,1^3)$ &  $ 40 \, \, T_3 T_1^3$\\
\hline
$(2^2,1^2)$ & $45 \, \, T_2^2 T_1^2$ \\
\hline
$(2,1^4)$ &  $-15 \, \, T_2 T_1^4$ \\
\hline
$(1^6)$ &  $ T_1^6$ \\
\hline
\end{tabular}
\end{center}
\end{table}
\\ The connection of these coefficients with the elementary symmetric polynomial $e_6(\lambda)$ is 
\beqr
p_6(\lambda)&=&(-1)^6 e_6(\lambda)=\sum_{1\leq i_1<\cdots<i_6\leq n}\lambda_{i_1}\cdots\lambda_{i_6} \non \\
&=& -\frac{1}{6!}\biggl(120\, T_6-144\, T_5 T_1-90\, T_4 T_2-40\, T_3^2+90\, T_4T_1^2+120\, T_3T_2T_1+15\, T_2^3 \non \\
&& -40\, T_3T_1^3 -45 \,T_2^2 T_1^2+15\, T_2 T_1^4-T_1^6\biggr)
\label{b3}
\feqr 
\section{Conclusions}
In this letter we present explicitly a diagrammatic way to calculate the coefficients of the characteristic polynomial of a square matrix. All information is encoded in combinatorial form into the sets of circle graphs constructed for all partitions of the index associated with the corresponding elemetary symmetric polynomial.
\vspace{1.cm}\\
{\bf Acknowledgements} \\
The author would like to thank Jiannis K. Pachos for valuable discussions and his comments on the manuscript.  
\bibliographystyle{plain}
 
\end{document}